# An Open Workflow Model for Improving Educational Video Design: Tools, Data, and Insights

Mohamed Tolba; Olivia Kendall; Daniel Tudball Smith; Alexander Gregg; Tony Vo; Scott Wordley
*Monash University,*
*Corresponding Author Email: Mohamed.tolba1@monash.edu*

## ABSTRACT

**CONTEXT**

Educational videos are widely used across various instructional models in higher education to support flexible and self-paced learning. However, student engagement with these videos varies significantly depending on how they are designed. While several studies have identified potential influencing factors, there remains a lack of scalable tools and open datasets to support large-scale, data-driven improvements in video design.

**GOAL**

This study aims to advance data-driven approaches to educational video design. Its core contributions include: (1) a workflow model for analysing educational videos; (2) an open-source implementation for extracting video metadata and features; (3) an accessible, community-driven database of video attributes; (4) a case study applying the approach to two engineering courses; and (5) an initial machine learning-based analysis to explore the relative influence of various video characteristics on student engagement.

**APPROACH**

This study introduces a structured, seven-stage workflow for large-scale analysis of educational video engagement. Implemented using open-source Python libraries (e.g., scikit-learn, OpenCV, Whisper), the workflow spans planning, data collection, dataset construction, exploratory analysis, modelling, insight extraction, and design feedback. All tools and documentation are publicly available and integrated into a web-based app, named *EduVideo Insights*, to support scalable analysis and benchmarking across diverse learning contexts.

**OUTCOMES**

The workflow enables scalable analysis of video features, such as duration, word count, speaking speed, and scene transitions, and their impact on viewer engagement (proxied by average percentage viewed). Initial results confirm known patterns (e.g., lower engagement with longer videos) and reveal unexpected trends, like higher word counts linking to increased retention. These findings offer early design insights while highlighting the need for richer datasets and more flexible models to guide effective video production.

**CONCLUSIONS**

Designing effective educational videos requires more than intuition. It calls for practical, data-driven methods that can inform and improve student engagement. This study offers a structured workflow model, open-source tools, and community-driven datasets, with early findings highlighting the relative impact of five key video features on audience engagement. By inviting contributions from educators, researchers, and policy makers, it lays the groundwork for a shared, evidence-based approach to educational video design.

# Introduction

"People learn more deeply from words and pictures than from words alone." (Mayer, 2005). Audiovisual media, particularly educational videos, have increasingly become a vital part of the teaching and learning process in tertiary education, driven by advances in technology and pedagogical strategies (Brame, 2017; Wordley et al., 2024). These resources offer students flexible, on-demand access to high-quality content, especially through open platforms such as YouTube. Video hosting platforms are generally classified as either *open* or *closed* (Giesen et al., 2022). Open platforms, such as YouTube and Dailymotion, allow unrestricted access to educational materials, supporting equitable learning opportunities at no cost. In contrast, closed platforms like Panopto, Echo360, and Kaltura limit access to authorised users, typically within institutional or subscription-based settings. Across both types, educational videos serve as a versatile tool across various instructional frameworks, whether in flipped classrooms, blended learning models, or fully online courses (Bishop & Verleger, 2013).

Despite their widespread adoption in higher education, which was further accelerated by the shift to online learning during the COVID-19 pandemic (Marinoni et al., 2020), the design of effective educational videos is not a trivial task. These videos encompass various characteristics and contextual factors, which were systematically classified by Navarrete et al., (2025) into eight categories: audio features, visual features, textual features, instructor behaviour, learners' activities, interactive features, production style, and instructional design. Each of these categories includes numerous attributes that should be considered during the video design process.

Numerous research efforts have investigated various characteristics of educational videos. One commonly explored feature is video length, which falls under the textual features category. Studies by Guo et al. (2014), Brame (2017) and Wordley et al. (2024) recommend keeping videos short (under six minutes) to maximise student engagement. In contrast, Lagerstrom et al. (2015) suggested that the maximum video length should be in the range of 12–20 minutes. Meanwhile, Giannakos et al., (2016) argue that learners may actually prefer much longer videos (close to 45 minutes), which could lead students to adopt videos more readily as part of their learning process. Another frequently examined characteristic is speech rate, which falls under the instructor behaviour category. It refers to how quickly the speaker delivers content in an educational video. Some researchers advocate for a fast speaking pace (around 185–254 words per minute) to sustain attention and convey enthusiasm (Guo et al., 2014; Brame, 2017). In contrast, van der Meij & van der Meij, (2013) recommend moderate speech rate to aid comprehension, especially when dealing with complex or unfamiliar content.

The current research landscape on educational video design reflects a fragmented body of knowledge, shaped by varying methodologies, learner populations, and instructional settings. As a result, recommendations often differ and can sometimes appear contradictory, although many remain valid within their specific contexts (Navarrete et al., 2025; Shoufan, 2019). While there is a growing volume of research examining the impact of individual features such as video length and speech rate on viewer engagement, most studies offer limited transferability and generalisability. A key gap remains in the availability of scalable tools, reproducible workflows, and open datasets to enable rigorous, systematic analysis across diverse learning environments.

This work addresses the identified gap by proposing a structured workflow model, an openly shared dataset, and an open-source implementation for the systematic analysis of educational videos. To demonstrate the practical utility of the approach, it is applied to a real-world case study involving lecture recordings from two undergraduate engineering courses hosted on YouTube. The primary aim is to identify which video features have the strongest influence on student engagement. The analysis focuses on characteristics such as video duration, number of words spoken, speaking speed, number of visual transitions, and visual transition rate. Engagement is proxied by the average percentage viewed, a metric provided by YouTube Analytics.

# Methodology

This study presents a seven-stage workflow model, grounded in the *Learning Analytics Cycle Framework* (Clow, 2012), for quantitatively analysing educational video design at scale. Developed for reproducibility and broad applicability, the approach is compatible with various video-hosting platforms and supports both educational and research contexts. The workflow is implemented in Python using open-source libraries such as *scikit-learn* and *numpy*, with all scripts, data, and documentation made publicly available via a [GitHub repository](). To further improve accessibility and promote community engagement, a web-based application, named [*EduVideo Insights*](), is also under development, enabling scalable analysis and benchmarking of educational videos. The proposed workflow model, outlined below, is designed to systematically collect, process, and analyse educational video data, transforming raw content into structured datasets that are ultimately interpreted using machine learning techniques.

*Stage 1: Planning and Video Selection*

- Define the purpose and context of the analysis.
- Select a consistent type of video content based on audience, pedagogical intent, and delivery format. For example, avoid combining lab demonstrations with lectures or workshops.

*Stage 2: Data Collection*

- Gather essential metadata, such as unique video identifiers, titles, author and upload date.
- Extract content features, such as video duration, speaking speed and visual transitions.
- Collect engagement metrics, such as average percentage viewed or likes vs dislikes.

*Stage 3: Dataset Construction*

- Integrate metadata, features, and engagement metrics using shared identifiers.
- Handle missing data, resolve inconsistencies, and standardise formats.
- Prepare a structured dataset ready for analysis and modelling.

*Stage 4: Exploratory Data Analysis*

- Visualise distributions, correlations, and trends across features.
- Assess data quality, patterns and feature variability.
- Identify candidate features for modelling.

*Stage 5: Model Training & Validation*

- Choose appropriate machine learning techniques based on data properties and goals.
- Train and validate models to assess feature influence on engagement metrics.

*Stage 6: Insight Extraction*

- Examine model outputs to determine the most influential predictors of retention.
- Simulate and identify how specific changes in feature values could affect predicted engagement outcomes.

*Stage 7: Educational Design Feedback*

- Translate analytical findings into design recommendations (e.g., video length, pacing, caption use).
- Provide actionable guidelines to improve future video-based instructional design.

# Case Study Data Selection

This paper applies the proposed workflow model to a specific case study aimed at exploring how specific video characteristics influence audience engagement and identifying which features have the most significant impact. The data used in this study comprises lecture videos from two engineering courses, *Engineering Materials* (90 videos) and *Modelling and Control* (54 videos), delivered at a regional Australian university during 2020 and 2021 by the same lecturer. These courses were selected primarily due to the authors' existing access to the lecture recordings and associated engagement data. Each course typically enrolled approximately 300 students and contributed to multiple engineering streams, including mechanical, mechatronics, electrical, aerospace, and medical engineering. Lecture videos were distributed via YouTube as unlisted playlists, with access provided to students through the institution's LMS, digital textbook, and communication channels. These videos were produced following recommended design practices proposed in (Guo et al., 2014), including segmentation and pacing.

# Results and Discussion

The implementation of the proposed workflow model on the presented case study was done through the web-based application we are currently developing, namely *EduVideo Insights.* Figure 1 shows the homepage of the application, with all workflow stages accessible via links on the navigation bar located on the left side of the page.

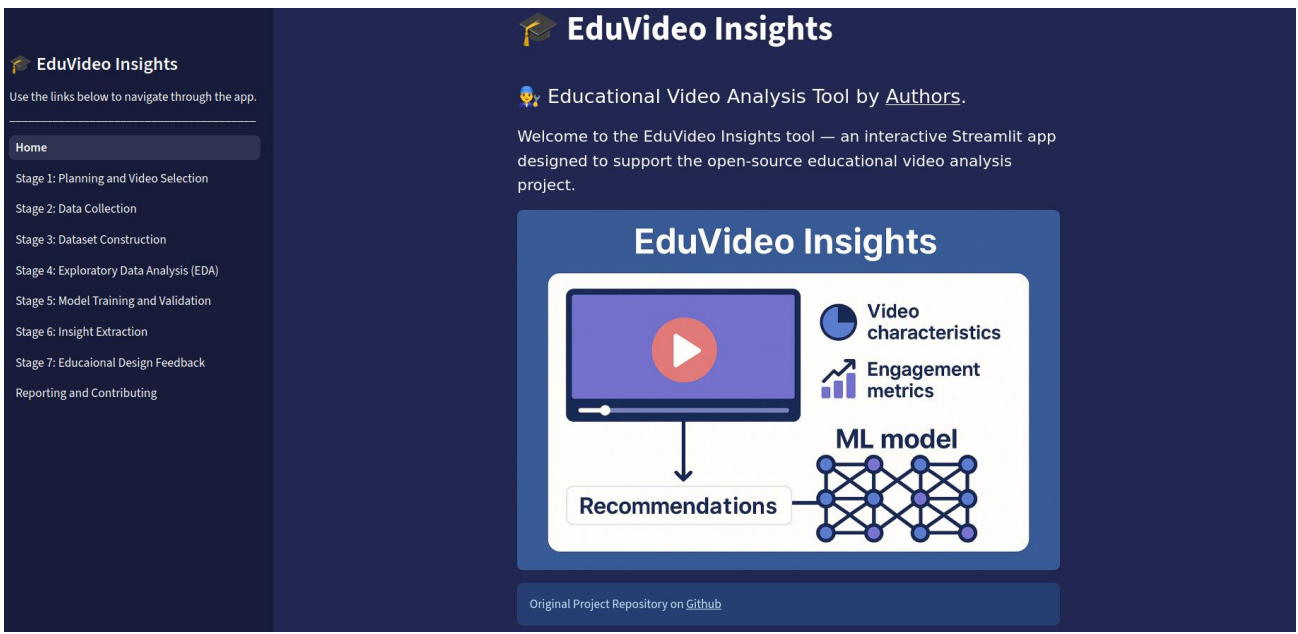

**Figure 1: Screenshot of the *EduVideo Insights* home page**

## Stage 1: Planning and Video Selection

The purpose of the analysis is to better understand how specific attributes of educational lecture videos, designed for undergraduate engineering students, affect student engagement (proxied by average percentage viewed). The attributes under investigation include video duration, total words spoken, speaking speed, number of visual scenes, and the rate at which these scenes occur. These were chosen due to their theoretical relevance to cognitive load and multimedia learning principles, and their potential influence on how students engage with educational video content (Brame, 2017).

**Stage 2: Data Collection**

Data collection involves video metadata, features, and engagement metrics. The retrieval and extraction of much of this information can be automated, enabling data collection at scale. Further details are provided below.

Since the videos used in this study are hosted on YouTube, metadata collection can be automated using the YouTube Data API, provided the video IDs are known. This process requires a valid API key, which can be created free of charge through the [Google Cloud Console](https://console.cloud.google.com). The *EduVideo Insights* application automates this metadata retrieval process, as shown in Figure 2. The extracted data includes video URL, title, date of publication, and the name and ID of the hosting channel. Additional data must be entered manually by the user, including institution name, speaker name, course code, course name, unit level, video type (e.g., Lecture.), and subject area. The application provides a user-friendly interface to help import and integrate all this information. Once all the metadata is populated, the application generates a unique *dataset tag*, which helps identify and manage datasets consistently.

**Figure 2: Screenshot of the *EduVideo Insights* metadata retrieval page**

Next, *EduVideo Insights* requires all lecture videos to be uploaded in order to automatically extract key video characteristics relevant to the study, as detailed below. Video duration is calculated using the OpenCV library by retrieving the total number of frames and the frame rate; dividing the number of frames by the frames per second yields the duration in seconds, which is then converted into minutes. Scene or slide transitions are also detected using OpenCV by comparing pixel-level differences between consecutive frames. If the difference exceeds a predefined threshold, a scene change is registered, and the total number of such transitions is recorded. The total word count is determined using the Whisper speech-to-text package, which transcribes the audio and enables word count extraction from the resulting transcript. With video duration, word count, and number of transitions available, the app derives additional features: average speaking speed in words per minute (wpm) and average scene change rate in scenes per minute (spm). These extracted features are then organised into a structured data file, linked to the corresponding video ID and dataset tag, as shown in Figure 3.

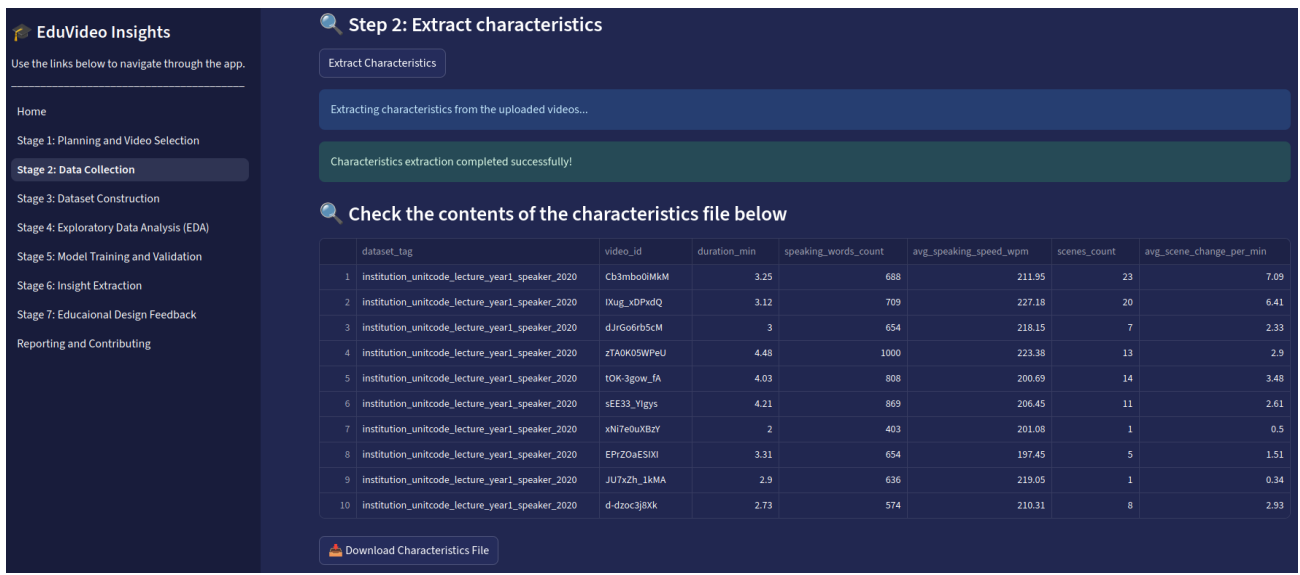

Figure 3: Screenshot of the *EduVideo Insights* characteristics extraction page

Finally, engagement metrics, specifically the average percentage viewed, are collected. As direct programmatic access to these metrics requires elevated permissions on the associated YouTube channel, *EduVideo Insights* does not currently support automated retrieval. Instead, we manually obtained this data via the YouTube Analytics dashboard and imported it into the application.

### Stage 3: Dataset Construction

During the dataset construction stage, metadata, extracted features, and engagement metrics were integrated using consistent identifiers, namely dataset tags and video IDs, to ensure accurate alignment across all data sources. This process involved addressing missing values, resolving inconsistencies, and applying standard formatting rules to ensure uniformity across the dataset. The resulting structured dataset was then prepared for downstream analysis and modelling, providing a reliable foundation for exploring the relationships between video characteristics and student engagement.

### Stage 4: Exploratory Data Analysis

As part of the exploratory data analysis, histograms were plotted for each extracted feature using 15 bins to examine their distributions and overall trends (see Figure 4). The distribution of video duration was slightly right-skewed, with the majority of videos ranging between 2 and 7 minutes. Similarly, word count showed a right-skewed distribution, with most videos containing between 500 and 1400 words. Average speaking speed followed an approximately normal distribution, clustered between 190 and 220 words per minute, indicating a relatively consistent delivery style across the dataset. Both scene count and average scene changes per minute were also right-skewed, with the majority of videos exhibiting fewer than 15 scenes and fewer than 4 scene changes per minute, respectively. Finally, the histogram of average percentage viewed revealed that most videos were watched between 60% and 90%, suggesting generally high levels of audience engagement. While the observed skewness indicates that the dataset may not represent an ideal distribution for studying the effects of all features uniformly, its availability and relevance to real-world instructional settings make it a valuable basis for extracting preliminary insights.

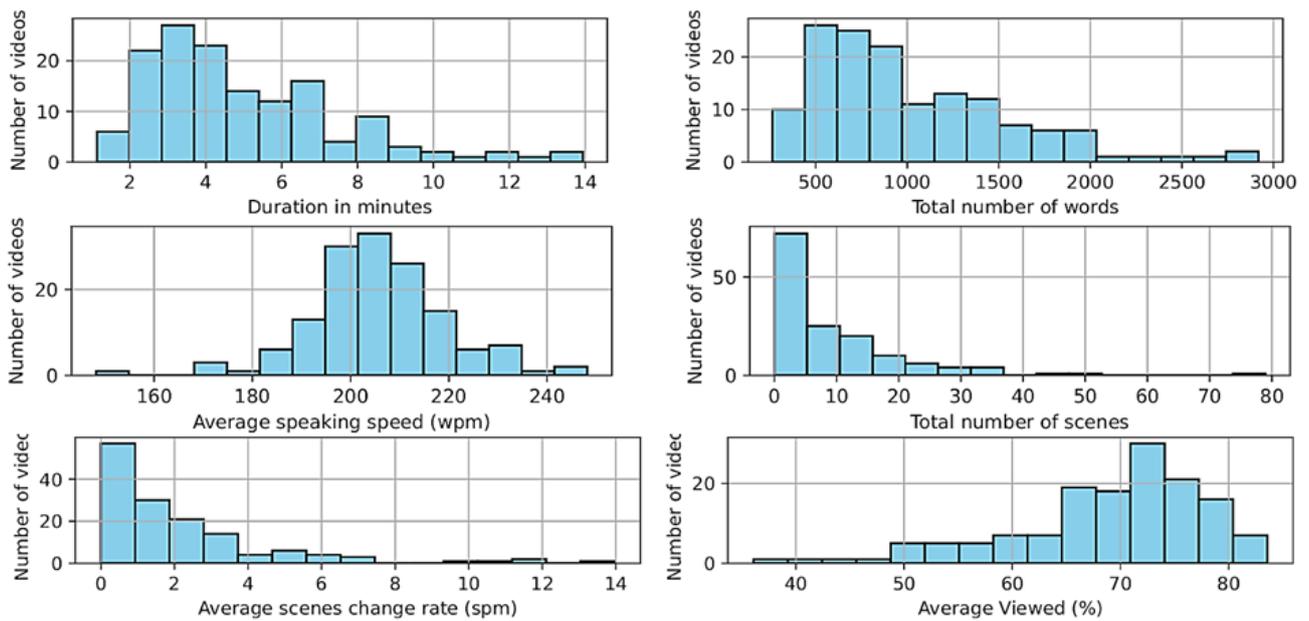

**Figure 4: Histograms of video features and engagement metric**

To help understand linear associations, a Pearson correlation analysis was conducted between each video feature and the average percentage viewed. As illustrated in Figure 5, video duration ($r = -0.23$) and word count ($r = -0.22$) showed the strongest negative correlations with engagement, suggesting that longer and more verbose videos are generally associated with lower viewer engagement. Average speaking speed and scene count demonstrated minimal correlation ($r = -0.03$ and $-0.01$, respectively), while average scene change rate revealed a slight positive correlation ($r = +0.09$), implying that moderately frequent visual transitions may have a small positive impact on engagement. Although none of the relationships were strong (all $|r| < 0.25$), the analysis provided initial insights into feature relevance and helped guide subsequent non-linear exploration.

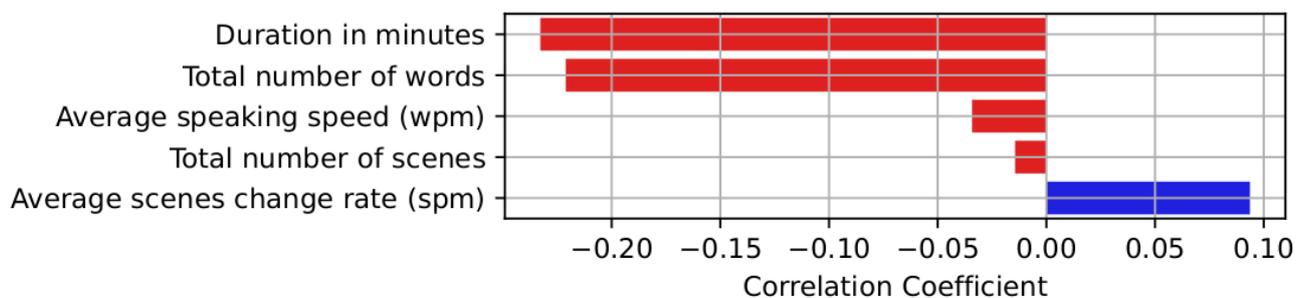

**Figure 5: Correlation of video features with average percentage viewed**

To further examine the relationship between video features and audience engagement, LOESS (Locally Estimated Scatterplot Smoothing) curves were plotted for each feature against the average percentage viewed. As illustrated in Figure 6, the smoothed trend lines reveal clear non-linear patterns, offering more detailed analysis than linear models alone. Shorter videos are consistently associated with higher percentage viewed, with engagement declining sharply as duration increases. Word count shows a similar pattern, where excessively lengthy transcripts correlate with reduced viewer attention. Speaking speed displays an optimal range around moderate rates, while both slower and faster delivery styles appear less engaging. For scene count and scene change rate, moderate values tend to correspond with higher retention, although this effect is comparatively weaker. These insights emphasise the value of balanced pacing and visual flow in the design of educational videos.

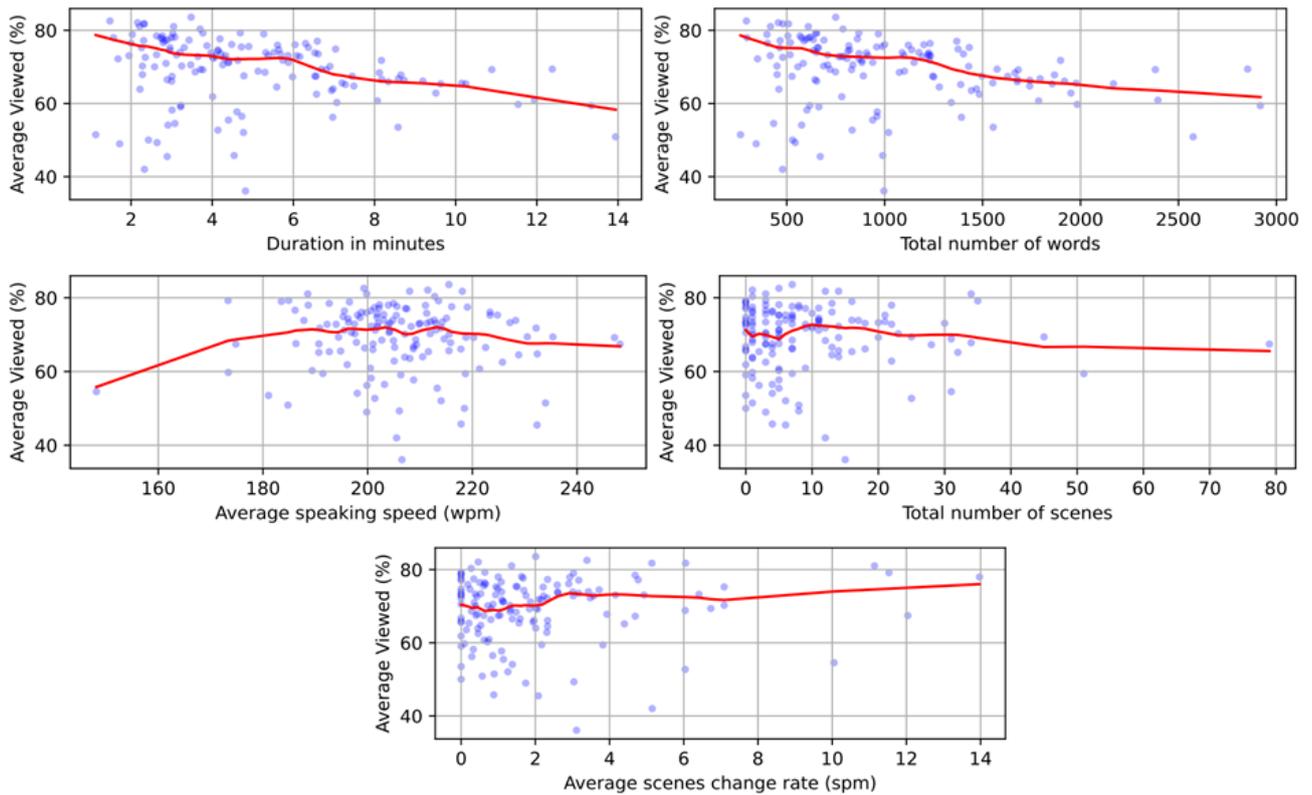

**Figure 6: LOESS curves of video features vs average percentage viewed**

The Pearson correlation and LOESS analyses suggest that while some features, such as video duration and word count, show modest linear relationships with engagement, others exhibit non-linear or weak associations. Based on these observations, features with clearer linear trends (e.g. duration and word count) were selected for modelling, while those with less defined patterns (e.g. speaking speed, scene count, and scene change rate) were included as part of a preliminary investigation into their potential influence on engagement.

### Stage 5: Model Training & Validation

A multiple linear regression model was selected for its simplicity and interpretability, serving as a baseline to evaluate how well the selected video features explain variation in the average percentage viewed. Prior to training, all features were normalised using z-score standardisation to achieve a mean of zero and a standard deviation of one, enhancing the interpretability of model coefficients by enabling direct comparison of their relative influence. The trained model yielded a root mean squared error (RMSE) of 8.6 and an $R^2$ value of 0.0853, indicating modest explanatory power. The resulting feature weights were –21.81 for video duration, 20.52 for total number of words, –3.04 for speaking speed, –1.02 for total scenes, and 1.15 for scene change rate. These coefficients reflect both the direction and relative strength of each feature's influence on predicted engagement. Figure 7 shows the predicted versus actual average percentage viewed, illustrating the model's limited yet observable alignment with real engagement outcomes.

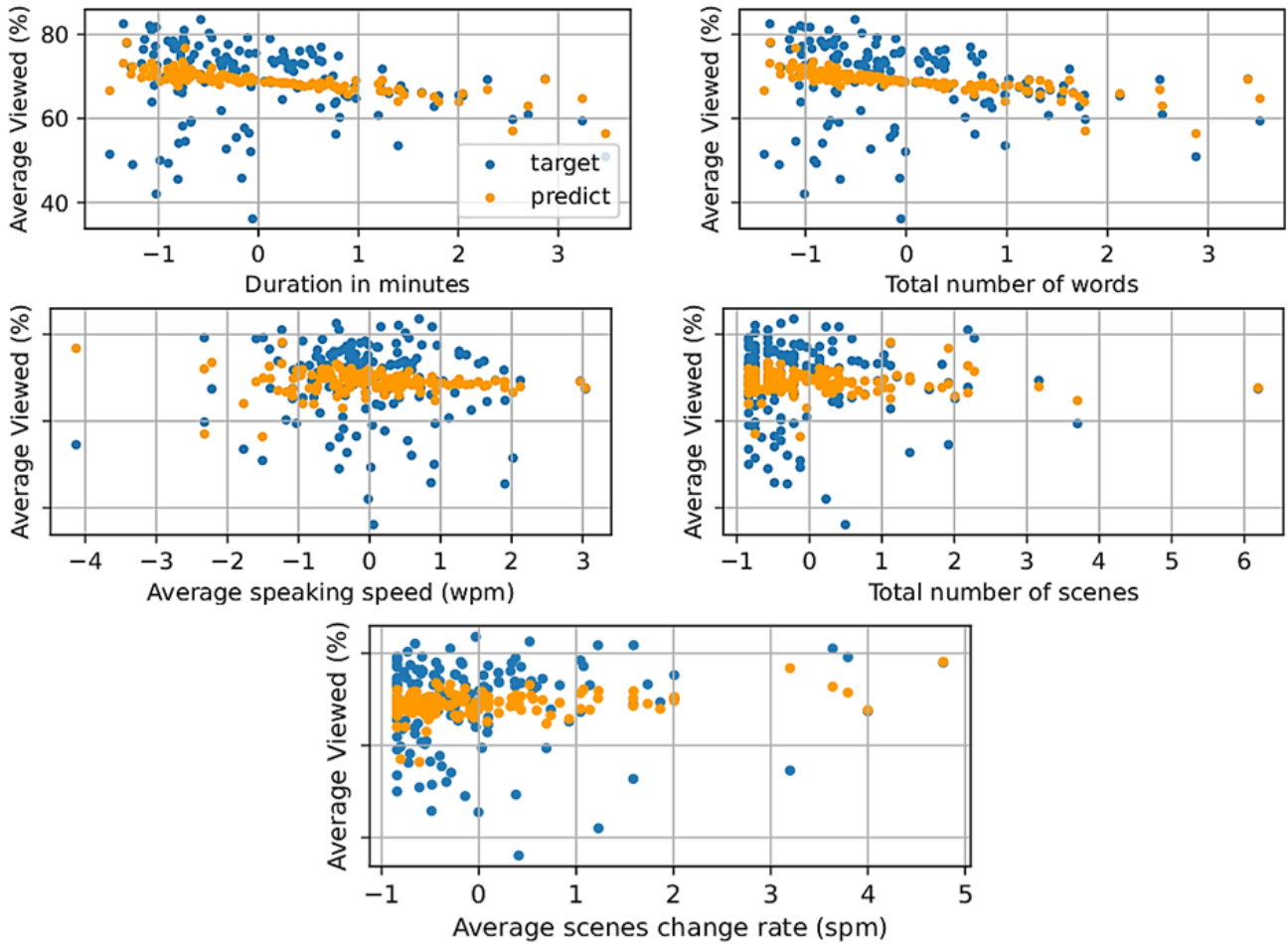

Figure 7: Predicted vs actual average percentage viewed (target) values using the trained model.

### Stage 6: Insight Extraction

The normalised weights produced by the trained model provide a basis for comparing the relative influence of video features on audience engagement. As expected, video duration showed a strong negative association with average percentage viewed and emerged as the most influential feature among those studied. However, contrary to initial assumptions and exploratory data analysis trends, where higher word counts were generally linked to lower retention, the total number of words exhibited a positive association with engagement. Other features demonstrated weak or inconsistent relationships. These findings suggest that more representative data and more flexible modelling approaches are needed to better account for potential non-linearity and interactions between features.

### Stage 7: Educational Design Feedback

While the analysis offers preliminary insights, the educational design implications should be interpreted with care. The negative influence of longer video durations reinforces existing guidance to keep videos concise, supporting learner engagement. However, the unexpected positive association between word count and engagement challenges prior assumptions and suggests that the relationship may be more complex than initially understood. This highlights the need for caution when translating quantitative findings directly into design recommendations. Until more representative data and advanced modelling approaches are incorporated, these results should serve primarily as exploratory indicators rather than definitive design rules. Nevertheless, the workflow demonstrates potential for supporting evidence-informed decision-making in video production, particularly as future iterations refine the underlying data and analysis.

## Conclusions and Future Work

This study introduced a seven-stage workflow model and an open-source tool for analysing educational video design at scale, integrating structured data extraction, engagement analysis, and interpretable modelling. Applied to lecture videos from two engineering courses, the workflow demonstrated its potential to support data-informed insights into video engagement. However, several limitations were identified. The multiple linear regression model exhibited limited predictive power, and some results such as the unexpected positive influence of word count contradicted trends observed in exploratory analysis. These inconsistencies may reflect the narrow scope of the dataset, and the limitations of linear modelling used.

Future work will focus on expanding the dataset, investigating more advanced modelling techniques, and incorporating qualitative data, such as instructional strategy or student feedback, to support a mixed-methods approach. Additionally, further investigation into established frameworks and theories (e.g., Mayer's Multimedia Learning Theory or KD4Ed) will help refine and inform the workflow model's structure and pedagogical relevance. To advance this work, we invite educators, researchers, and institutions to contribute video data and metadata. Through open collaboration, the community can collectively improve the dataset, strengthen the methodology, and move toward robust better-informed guidelines for effective educational video production.

## Copyright statement